# Novel non-local behaviour of quasi-3D Wide Quantum Wells

J. Oswald, G.Span

Institute of Physics, University of Leoben, Franz Josef Str. 18, A-8700 Leoben, Austria

*We investigate the high magnetic field regime of wide quantum wells (WQW) for the case of a many valley host semiconductor. The complete system is described within a modified Landauer-Büttiker formalism and we demonstrate that a parallel contribution of two electron systems in different valleys of the band structure can lead to an edge channel related non-local behaviour even in the 3D-regime. From the obtained general result we derive also a simplified model which applies for the case of much different dissipation. It represents the most dissipative system by an Ohmic resistor network and the less dissipative system by an EC-system.*

The formation of QH-plateaux is well understood by the existence of dissipation less edge channel (EC) transport which appears due to the absence of back scattering directly in the EC's at the same edge [1]. On the basis of a Landauer-type formulation also the finite resistance in the regime between QH-plateaux can be understood in terms of EC back scattering across the bulk region of the conductor [2]. This back scattering is possible if the Fermi level is near the centre of a broadened Landau level (LL) and does not happen if the Fermi level is in a DOS-gap between LL's. Therefore basically the realization of a 2D-system is required for the observation of the QHE. The question if it is possible or not to observe the QHE also in a 3D electron system has been investigated by Störmer et al. who demonstrated that the QHE can be observed in a superlattice (SL) with a well pronounced 3D-behaviour as well[3]. Although there are no DOS- gaps between the LL's of the individual mini band states of a SL, the QH-plateaux appear if the Fermi level is in a mini band gap. Therefore one can conclude that the 3D-behaviour does not directly destroy the EC-transport but in order to avoid dissipation there must not be any coupling between the EC's at opposite edges. Consequently EC-transport should play a role also in the quasi-3D regime of a wide quantum well (WQW). In contrast to the SL-system there will be a permanent coupling between opposite EC's and hence also a permanent dissipation will occur. On this background a WQW can be understood to be a system which allows to study a gradual "turning on" of the 3D-behaviour.

The realization of wide quantum wells with a flat potential in the electron channel requires a parabolic bare potential which is difficult to obtain in GaAs/Al$_x$Ga$_{1-x}$As hetero structures by MBE growth [4]. The use of the host semiconductor PbTe allows to create WQW's without the necessity of remote doping and therefore no technological limit towards true 3D-samples exists. Therefore our model system is based on the boundary conditions which are known from already realized PbTe-WQW's [5]. The main facts are listed in the following:

(i) Due to the high electron mobility ($\mu > 10^5$ cm$^2$V$^{-1}$s$^{-1}$) it is easy to achieve $\mu \cdot B \gg 1$ already at low magnetic fields.

(ii) Because of the many valley character of PbTe there are two different effective masses for both, the subband splitting $\Delta E$ and the LL-splitting $\hbar\omega_c$. $\Delta E$ for the small mass is of the order of 1meV and about an order of magnitude smaller for the large mass. Therefore we are in the regime of $\hbar\omega_c > \Delta E$ already at low magnetic fields.

(iii) Also the condition $\hbar\omega_c > E_F$ is achieved already at moderate magnetic fields. Due to the high filling factor the Fermi level $E_F$ remains in the upper subbands even at high magnetic fields ($\Delta E < E_F$). This leads finally to a situation which is opposite to a standard 2D-system: At high magnetic fields where the Fermi level is already down in the lowest LL of both electron systems a large number of subbands can still be occupied.

(iv) Due to the large dielectric constant of PbTe ($\varepsilon \approx 1000$) the amplitude of native potential fluctuations is of the order of 1meV and thus of comparable magnitude as the subband splitting $\Delta E$.

(v) The mean distance of dopants in the electron channel is between 10 and 20 nm which results in a lateral size of the potential fluctuations of the order of 100nm. This is approximately a factor of 10 larger than the cyclotron radius at usual magnetic fields of 10 Tesla.

(vi) There is found an overall linear increase of the magneto resistance R$_{xx}$ with magnetic field in almost all epitaxially grown high mobility PbTe layers. This linear magneto resistance has been also the subject of investigations in other systems [6]. An explanation by Büttiker [7] is based on EC-conduction in the presence of lateral inhomogeneities in a quasi-3D electron channel.

(vii) Conductance fluctuations in the magneto conductance of macroscopic PbTe WQW samples have recently been observed [8]. The amplitude was found to be close to e$^2$/h which cannot be explained by diffusive bulk transport. The amplitude suggests a contribution of 1D-channels for which EC-conduction seems to be a plausible explanation.

(viii) Finally a well pronounced non-local behaviour has been observed experimentally in PbTe WQW's [8].

An explanation for the non-local effects is given in [9] and it is based on an equivalent circuit combining EC and bulk transport. Therefore the intention of this paper is to give a well defined theoretical basis for the application of these equivalent circuits.

From the facts listed in Pos. i-viii we conclude that the regime of EC-transport is already achieved in PbTe WQW's. The native potential fluctuations are of the order of the subband splitting and therefore magnetic bound states ("EC-loops") can be created in the bulk region at any magnetic field. Therefore a permanent coupling between the opposite edges across a system of coupled EC-loops in the bulk region is possible. We start the modeling of the high magnetic field regime on the basis of a Landauer type formulation and consider first a single valley situation. In contrast to other calculations [2] we use a back scattering probability $p$ for edge electrons per unit of length, which accounts for the fact that there are no discrete back scattering barriers in a quasi 3D sample.

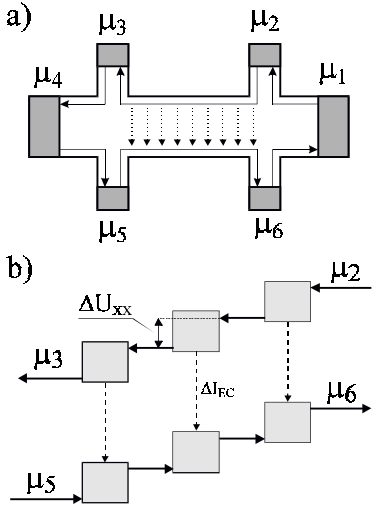

Fig.1: a) Basic scheme of a four terminal measurement. The edge channels (EC) are represented by the arrows connecting the contact reservoirs. The dashed arrows indicate the EC back scattering which leads to dissipation. b) Scheme for the modeling in the presence of permanent EC back scattering which happens continuously along the sample. The sample is divided into finite elements (shaded boxes) in which the scattering between EC's (horizontal arrows) and the EC-loops in the bulk occurs. The associated net "edge current" across the bulk region $\Delta I_{EC}$ is represented by the dashed arrows.

Fig.1 shows schematically how this concept works. In order to calculate the longitudinal voltage drop $\Delta U_{xx}$ in one discrete cell, one has to consider the edge current balance in this cell. On one hand one gets the net edge current of the sample by $I_{EC} = \nu \cdot (e^2/h) \cdot U_H$ with the Hall voltage $U_H = (\mu_2 - \mu_6) = (\mu_3 - \mu_5)$ and the filling factor $\nu$. On the other hand we have an "edge current" across the bulk region $\Delta I_{EC}$ which is associated with the back scattering process. $\Delta I_{EC}$ appears formally as a "loss" of edge current from the discrete cells at the upper edge according to $\Delta I_{EC} = p \cdot \Delta L \cdot I_{EC}$ where $\Delta L$ is the length of the cell and $p$ is the transmission probability for an edge electron to the opposite edge per unit of length. According to the EC-picture, this "loss" of current $\Delta I_{EC}$ from the cell must appear as a potential difference $\Delta U_{xx}$ between the incoming and leaving EC's at the same edge according to $\Delta I_{EC} = \nu \cdot (e^2/h) \cdot \Delta U_{xx}$. At the lower edge $\Delta I_{EC}$ enters the cells and hence the edge current balance leads to the opposite effect. In this way the potential difference between upper and lower EC's does not change because both EC-potentials are shifted in parallel by $\Delta U_{xx}$ (Fig.1b). In agreement with the general result of the Landauer-Büttiker formalism this means that the edge current in the longitudinal direction of the sample remains unchanged by the back scattering process. In this way we find a relation between the Hall voltage $U_H$ and the longitudinal voltage drop $\Delta U_{xx}$. For a homogeneous conductor (despite the potential fluctuations) with distance L between the voltage probes we get consequently:

$$U_{xx} = p \cdot L \cdot U_H \qquad (1)$$

In WQW's we have a small subband splitting which does not change much with the magnetic field B. Therefore also the number of EC-loops, which result from the potential fluctuations in the bulk region, does not change much. This should finally result in an almost constant coupling between the opposite edges and we can expect an overall linear increase of $R_{xx}$ with the magnetic field. However, as shown in [5] the level sequence is mainly determined by a square well confinement. Since the density of the EC-loops will depend on the level splitting at the Fermienergy, some oscillatory behaviour of p as a function of magnetic field can be expected. This leads basically to the bulk Shubnikov de Haas oscillations, which appear superimposed on the linear slope of $R_{xx}$. This effect is the subject of present work and is considered for publication else where.

We turn now to the many valley situation and investigate a parallel contribution of two electron systems in different valleys of the band structure. The further treatment is based on the assumption that intervalley scattering is sufficiently suppressed and a mixing of both systems is possible at the metallic contacts only. The suppression of intervalley scattering is addressed once more at the end of this paper. We would like to point out that the following treatment can be applied also any other system of parallel conducting layers in the quantum Hall regime.

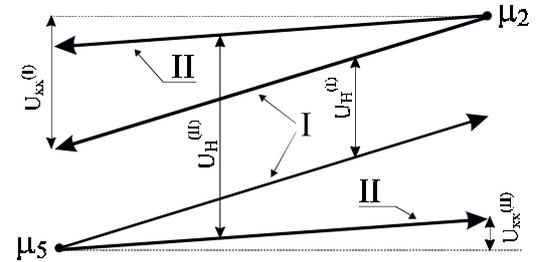

Fig.2 Combined potential scheme of two independent electron systems between the voltage probes for a sample structure like Fig.1a. The discrete longitudinal "potential stair case" of Fig.1b is replaced by a continuous slope which results from shrinking the cells to infinite small size. Due to dissipation, a characteristic longitudinal slope of the EC-potential appears for both electron systems.

Both systems I and II are characterised by the factors $p_I$ and $p_{II}$ which will be in general different. If we assume that $p_I > p_{II}$, the longitudinal slope of the EC's in system I will be more steep than for system II. As shown in Fig.2 the EC's of both systems start with the same potential at contact 2 and at contact 5 but the potentials of the EC's will not match if arriving at contact 3 and contact 6. As evident from Fig.2 the only known potential difference so far is the two-terminal voltage $U_{2P}$ which is given by $U_{2P} = (\mu_2 - \mu_5)$. $U_{2P}$ is the same for both electron systems and can therefore be written in two ways: $U_{2P} = U_H^{(I)} + U_{xx}^{(I)}$ or $U_{2P} = U_H^{(II)} + U_{xx}^{(II)}$. The superscripts (I) and (II) indicate the associated electron systems. According to Eqn. 1 we have $U_{xx} = L \cdot p \cdot U_H$ in each system and therefore we can rewrite the two representations of $U_{2P}$:

$$U_{2P} = U_H^{(I)} \cdot (1 + L \cdot p_I) = U_H^{(II)} \cdot (1 + L \cdot p_{II}) \quad (2)$$

The total current $I_{tot}$ can be written in terms of the individual Hall voltages

$$I_{tot} = \nu_I \frac{e^2}{h} U_H^{(I)} + \nu_{II} \frac{e^2}{h} U_H^{(II)} \quad (3)$$

where $\nu_I$ and $\nu_{II}$ are the individual filling factors. According to Eqn. 2 we use $U_H = U_{2P}(1 + p \cdot L)^{-1}$ for both systems and substitute the Hall voltages in Eqn. 3 by $U_{2P}$

$$I_{tot} = U_{2P} \frac{e^2}{h} \left[ \frac{\nu_I}{1 + L \cdot p_I} + \frac{\nu_{II}}{1 + L \cdot p_{II}} \right] \quad (4)$$

For this general result we can consider some special cases: (i) If $L \cdot p \ll 1$ is valid for both systems, we have the normal QH-situation where $U_{2P}$ is the pure Hall voltage which is the same for both systems. The total current is divided according to the individual filling factors. (ii) If $L \cdot p \gg 1$ for both systems, the current is divided according to the factors $\nu \cdot (L \cdot p)^{-1}$. This is formally a simple parallel conduction of Ohmic two-terminal resistors. (iii) If $p_I \gg p_{II}$ it is always possible to choose the distance L according to $p_I^{-1} < L < p_{II}^{-1}$ so that we have $L \cdot p_I > 1$ and $L \cdot p_{II} < 1$ at the same time. This can be expected to be possible for e.g. the PbTe-WQW's because of the significantly different subband splitting of both electron systems. The most simple situation appears if $L \cdot p_I \gg 1$ because for this case it is a sufficient approximation to use an Ohmic (two-terminal) resistor for representing the most dissipative system I. If we have also $L \cdot p_{II} \ll 1$ at the same time, it is a sufficient approximation to represent the less dissipative system II by pure EC-conduction. In this case we get $U_{2P} = U_{xx}^{(I)} = U_H^{(II)}$ from Eqn. 2 which finally allows to simplify Eqn. 4 and we get:

$$I_{tot} = U_{xx} \left( R_{xx}^{-1} + \nu_{II} \frac{e^2}{h} \right) \quad (5)$$

$R_{xx}$ represents the bulk resistance due to the most dissipative system I. Based on Eqn. 5 it is possible to describe the behaviour of a many valley system with significantly different dissipation in the involved valleys formally by a parallel contribution of an Ohmic bulk system and an EC-system. The most important advantage of this simplification is that it allows to model also a complicated sample geometry with a finite bulk resistance of the voltage probes (see lower insert of Fig.3). For modelling one must consider, that in contrast to a standard QH-system all contacts will be usually at different potential. Therefore a detailed current balance for each contact of the network is required. As an example, at contact 3 (insert Fig.3) the entering EC's will be at potential $\mu_2$ but the leaving EC's will carry $\mu_3$. This implies a net edge current to the reservoir of contact 3. In order to keep the current balance, a returning "bulk" current through the most dissipative system must appear in the voltage probe of contact 3. Basically one has to calculate the following current balance at each contact i:

$$I_i^I = \nu_{II} \frac{e^2}{h} (\mu_{i-1} - \mu_i) \quad (6)$$

$I_i^I$ is the "bulk" current which leaves contact $i$ through the dissipative system I. This current compensates the net edge current of system II to contact $i$ which is represented by the right part of Eqn. 6. A solution for the whole network can be found numerically by applying Eqn. 6 to each contact: First one starts with a pure "bulk solution" for the network (solution without EC's) and calculate the potentials at the contacts. From this potentials the bulk current through the voltage probes can be found by applying Eqn. 6. This allows to re-calculate the current distribution in the resistor network. From this, a new set of potentials is obtained from the voltage drop of the bulk current at the individual resistors. This leads again to new edge currents and so on. A solution for the whole network is found, if the potentials obtained from the voltage drop in the bulk system lead to edge currents which are already in balance with the bulk current.

In order to demonstrate one particular application of this model, we present a model calculation of a Hall bar with 3 pairs of Hall probes with finite bulk resistance (Fig.3). In this calculation the influence of a finite bulk resistance of the "unused" Hall contact pair 3-7 on $R_{xx}$ between the outer voltage probes 4-2 ($R_{xx}=R_{42}$) is demonstrated. The magnetic field is kept constant at 5 Tesla and the number of EC's was set to $\nu = 6$. The bulk resistance of the voltage probes of contact 3 and 7 has been varied simultaneously between zero and 10 k$\Omega$ while the values of the other Ohmic resistors of the network have been kept constant at values which can be obtained from Fig.3. Classically the pure existence of the middle Hall contact pair with open contacts should have no influence on $R_{42}$. But one can clearly see that $R_{42}$ decreases as $R_3$ and $R_7$ increase. From the choice of a high bulk resistance for the main path (each resistor 50kΩ) one can conclude that the Hall bar is close to a standard QH-situation. However, the comparatively low bulk resistance between the opposite contacts of the Hall probes provide an additional possibility for EC-back scattering. Therefore we see basically a Landauer-type behaviour of $R_{24}$ as a function of $R_3$ and $R_7$. The bulk current $I_3$ which leaves contact 3 is nearly equal to the total current if $R_3 = R_7 = 0$ (upper insert of Fig.3). However, this current "disappears" again into the opposite contact 7 ($I_7 = -I_3$) and therefore does not contribute to the sample current which remains to be mainly an edge current.

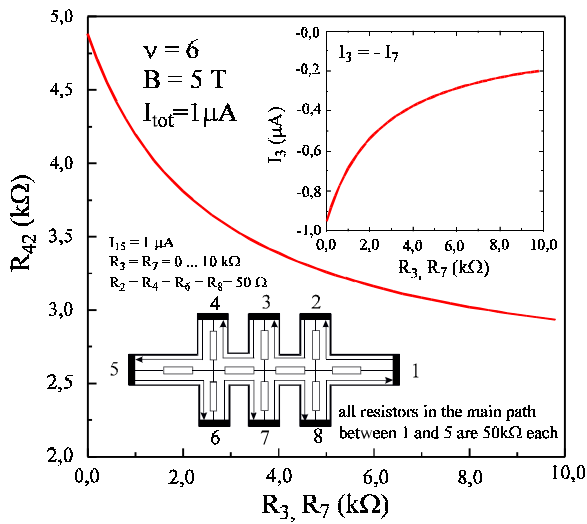

*Fig.3 Result of a model calculation for a Hall bar with 3 Hall contact pairs with finite bulk resistance of the voltage probes.*

Finally we would like to compare our model system to a standard single valley QH-system. In a standard QH-system a signal between non-local voltage probes appears only at magnetic fields between QH-plateaux [10]. In this regime the states of the upper LL become extended into the bulk region and are almost de-coupled from the EC's of the lower LL's. Therefore we have an independent bulk- and EC-system even in the standard single valley situation which is similar to the situation discussed above. However, in order to get such a well defined splitting of a single valley electron system into an independent bulk- and EC-system, well developed DOS-gaps between the LL's are required. Therefore in standard single valley systems the observation of non-local EC-transport must be closely related to the possibility of observing also the QHE. In WQW's the appearing 3D-behaviour does not allow such well separated systems within a single valley. This is evident from the absence of the QHE and the permanently present dissipation. At this point the existence of a second electron system with different dissipation in a different valley is necessary for the non-local behaviour which now can appear also without the QHE. A major point is that a scattering between the different electron systems is sufficiently suppressed outside the metallic contacts. In PbTe-WQW's a scattering between the four [111] L-point valleys is basically forbidden because of symmetry reasons. However, a defect assisted scattering between the valleys is in principle still possible. Up to now it is not absolutely clear how defect assisted intervalley scattering could be suppressed in real structures. So far there can be found one strong argument which is related to an effect, which is also responsible for the suppression of back scattering of edge electrons in EC's at the same edge. Recent calculations by Müller [11] demonstrate explicitly how the wave functions of the edge electrons manage to avoid back scattering in the presence of a disorder potential. It was shown that in a magnetic field the wavefunction of an edge electron gives rise to an excluded area around the scattering centre. From this effect one can conclude that in a sufficient magnetic field the electron system is able to incorporate the disorder potential into the solution for the wavefunction without destroying the well defined macroscopic behaviour of the EC. Consequently the disorder potential should not act as a scattering centre anymore. But in this case any type of defect assisted scattering should be suppressed (including defect assisted intervalley scattering).

In summary we have investigated theoretically the high magnetic field regime of high mobility wide quantum wells. We have shown that in the case of a many valley host semiconductor edge channel related non-local effects can persist far into the 3D-regime where the QHE is already quenched in single valley systems. In addition we have presented a procedure for solving complex equivalent circuits which can be applied in principle to any QH-system which contains parallel conduction.

Financial support by Fonds zur Förderung der wissenschaftlichen Forschung Vienna (Project: P10510 NAW)